# Non-Hermitian Dynamics in Quantum Anomalous Hall Insulators

Le Yi[1], Emma Steinebronn[1], Asmaul Smitha Rashid[2], Nitin Samarth[1,3], Ramy El-Ganainy[4], Şahin L. Özdemir[4], and Morteza Kayyalha[2, *]

[1]Department of Physics, The Pennsylvania State University, University Park, Pennsylvania 16802, USA
[2]Department of Electrical Engineering, The Pennsylvania State University, University Park, Pennsylvania 16802, USA
[3]Department of Materials Science and Engineering, The Pennsylvania State University, University Park, Pennsylvania 16802, USA
[4]Department of Electrical and Computer Engineering, Saint Louis University, St. Louis, Missouri 63103, USA

*Corresponding author: mzk463@psu.edu

Magnetically doped topological insulators (TIs) exhibit two distinct phases: the quantum anomalous Hall (QAH) phase when the Fermi level resides within the surface gap, and a metallic phase outside the gap. The QAH phase hosts unidirectional transport channels known as chiral edge states, while the metallic phase exhibits non-reciprocal transport due to unbalanced bidirectional edge states. Utilizing the chiral edge states in Cr-doped (Bi, Sb)$_2$Te$_3$ sandwich structures, we realize non-Hermitian conductance matrices in a one-dimensional Corbino chain with well-defined chirality. By tuning the boundary conditions from open to periodic, we reveal the non-Hermitian skin effect, where eigenstates localize exponentially at one end of the chain. In the metallic phase, we further observe asymmetric, bidirectional coupling between the neighboring sites in the conductance matrix, a direct consequence of the system's intrinsic non-reciprocity. These results establish magnetic TIs as a powerful platform for investigating emergent non-Hermitian phenomena in topological systems.

**INTRODUCTION**

Recent advances in non-Hermitian physics have provided a rigorous framework for analyzing dissipative quantum systems, where gains and losses lead to complex energy spectra and violation of energy conservation (*1-3*). Non-Hermitian systems and their hallmark signatures known as exceptional points (EPs), spectral degeneracies where both the eigenvalues and the associated eigenstates of the system coalesce, have been studied both theoretically and experimentally in diverse platforms; these include photonic systems (*4-17*), atomic gases (*18-22*), electrical circuits (*23-32*), magnonic systems (*33-35*), and quantum systems (*36-39*). These platforms have enabled applications spanning precision sensing (*40-43, 29, 44-46*) and TI lasers (*47, 48*). A key manifestation of non-Hermiticity is the non-Hermitian skin effect (NHSE) (*49*), which is characterized by the exponential localization of eigenstates at the system boundaries. Unlike Hermitian Hamiltonians, which satisfy $H\Psi = E\Psi$ with real eigenvalues, non-Hermitian Hamiltonians allow for complex eigenvalues, $H\Psi = (E + i\alpha)\Psi$, where α is a real number that captures the dissipative effects (*3*). The connection between non-Hermitian topology and Hermitian topological quantum materials has recently been explored (*50-52*). In this work, we investigate the interplay between non-Hermitian dynamics (*53, 50, 51*) and non-reciprocal transport in a quantum anomalous Hall (QAH) insulator (*54*) as depicted by the schematics in Fig. 1. The QAH insulator utilized in our study is a magnetically doped Cr-(Bi,Sb)$_2$Te$_3$ (*55-58*) that exhibits two distinct phases, namely, the QAH phase at the charge neutrality point (CNP), and a metallic phase away from the CNP. In the QAH phase, time-reversal symmetry is explicitly broken by magnetic doping, leading to a gap opening in the Dirac surface states and the emergence of a single chiral edge state in the absence of an external magnetic field. The chiral edge state facilitates unidirectional, dissipationless electrical transport along the boundary of the TI (*59, 55*). In this work, we utilize this robust chiral edge of magnetic TIs in a Corbino geometry to introduce non-Hermiticity (*50*). We show that in the metallic phase, the interplay between the Dirac surface states and the magnetization, and the resulting non-reciprocal behavior lead to bidirectional coupling and the non-Hermitian skin effect in the Corbino geometry. Furthermore, we implement the Hatano-Nelson (HN) model in the QAH phase of the magnetic TI, where the contact arms in the Corbino geometry (as shown in Fig. 1B) play the role of lattice sites in the HN model. The HN model is a one-dimensional paradigmatic non-Hermitian lattice model that incorporates asymmetric hopping amplitudes, explicitly breaking the Hermiticity, and leading to the non-Hermitian skin effect (NHSE) (*60*). Within the context of tight binding models (or coupled mode theory in optics), the HN Hamiltonian is given by:

$$\mathcal{H}_{\text{HN}} = \sum_i J_{\text{left}} c^\dagger_{i-1} c_i + J_{\text{right}} c^\dagger_{i+1} c_i = \boldsymbol{c}^\dagger H_{\text{HN}} \boldsymbol{c}, \quad (1)$$

where $J_{\text{left}}$ and $J_{\text{right}}$ are the left and right hopping amplitudes between adjacent sites in the one-dimensional HN lattice, respectively, and $J_{\text{left}} \neq J_{\text{right}}$ induces a preferential accumulation of the eigenstates at one boundary, thereby breaking the conventional bulk-boundary correspondence (*61-64*). In Eq. (1), $c_i$ and $c_i^\dagger$ are the annihilation and creation operators for the lattice site $i$, respectively. In our studies, we observe that by tuning the Fermi level of our TI, via a back gate, we can continuously move the system from the QAH phase to the metallic phase, allowing *in-situ* control of $J_{\text{left}}$ and $J_{\text{right}}$. A key advantage of our approach is that it functions without requiring a constant external magnetic field, thereby expanding its applicability to a wide range of devices. This is particularly important for non-Hermitian sensing, where magnetic fields can hinder the sensitivity of the device (*65-67*).

**RESULTS**

We investigate non-Hermitian dynamics in multi-terminal magnetic TI devices made in a Corbino geometry. The magnetic TI is a sandwich of 3 quintuple layer (QL) Cr-doped (Bi, Sb)$_2$Te$_3$ / 5 QL (Bi, Sb)$_2$Te$_3$ / 3 QL Cr-doped (Bi, Sb)$_2$Te$_3$ sitting on top of a SrTiO$_3$ (STO) substrate; see method section for more details about material growth. Figure 2A shows an optical image of a representative device, along with its cross-sectional schematic. Each contact arm of the device has an inner and outer electrode. We begin by investigating the magneto-transport properties of the device at temperature $T = 100$ mK. An AC excitation current ($I_{\text{ac}} = 1$ nA) is injected into the inner arm using a lock-in amplifier as shown in Fig. 2A. The magnetic field is then swept within the ±0.5 T range at a back gate voltage of $V_g = V_g^0 = 4$ V, where $V_g^0$ corresponds to the CNP. The resulting magnetic field dependences of the longitudinal ($R_{xx}$) and Hall ($R_{xy}$) resistances, normalized with respect to the resistance quantum, $h/e^2$, are shown in Fig. 2B. We observe a hysteresis loop characterized by a quantized value (0.99 $h/e^2$) of $R_{xy}$ accompanied by vanishing (0.015 $h/e^2$) $R_{xx}$. This observation is a characteristic hallmark of the QAH effect (*55*). The deviation from perfect quantization in $R_{xy}$ primarily arises from the lack of precise calibration of the lock-in amplifier and the resistors used for high-accuracy measurements, whereas the nonzero $R_{xx}$ indicates an imperfect QAH phase. For the proceeding measurements, we first magnetize the device at $B = +0.5$ T and then set the magnetic field to $B = 0$ T.

To explore the NHSE, we use spectral reconstruction whereby the conductance matrices, $G$, are retrieved from a series of electrical measurements. This is achieved by first measuring the resistance matrices, $R$, and obtaining $G$ from $G = R^{-1}$. The experimental setup for measuring a particular $R$ matrix element, $R_{32}$, is shown in Fig. 3A. Contact arms labeled 1-5 are used for measurements. The inner electrode of each arm is employed for current injection and voltage measurement, while the outer electrode is grounded. The current through each arm is monitored via the outer electrode using a current-to-voltage preamplifier, which provides a virtual ground. The contact arm labeled BC is used for tuning the boundary condition of the HN lattice. To measure $R_{32}$, we inject $I_{ac} = 1$ nA into the inner electrode of the contact arm 2 using a lock-in amplifier, and measure the voltage drop, $V_3$, across the inner electrode of the contact arm 3 and ground. Due to the presence of multiple grounds, current will be distributed across the device, and the distributed current will flow from the device to the measurement setup through the fridge lines, each having a resistance of $R_{fridge} = 2.93$ k$\Omega$. To obtain an accurate estimate for $R_{32}$, the voltage dop across the fridge line needs to be subtracted from $V_3$ (see method section). A variable resistor, $R_{BC}$, is also connected between the inner electrode of the contact arm BC to the ground (see Fig. 3A). By tuning the value of $R_{BC}$, we can continuously tune the boundary condition of the finite lattice from OBC ($R_{BC} = 0$) to periodic boundary condition, PBC ($R_{BC} = \infty$). However, in our cryostat, $R_{fridge} = 2.95$ k$\Omega$ limits the lowest value of $R_{BC} = R_{fridge}$. To distinguish this scenario from the true OBC with $R_{BC} = 0$, in the following, we refer to the smallest $R_{BC} = R_{fridge}$ as the pseudo-OBC.

By systematically permuting the injected current and voltage probes along the labeled arms, we obtain a complete set of measurements that allow us to reconstruct each column of the matrix $R$. Figure 3B shows the colormap of the resulting $G = R^{-1}$ matrix, with the matrix elements normalized with respect to the conductance quantum, $e^2/h$ (see Fig. S1 for values of the matrix elements in the SI). The diagonal and the lower bidiagonal elements of G are populated with values of 2 and -1, respectively. In the case of pseudo-OBC, the matrix element $G_{15}$, which corresponds to the coupling strength between the first and the last sites of the chain, is -0.11, which is different from $G_{15} = 0$ in the ideal OBC case with $R_{BC} = 0$, as can be obtained from a typical Landauer-Büttiker scattering matrix for a five-terminal mesoscopic setup (*68*). However, the slightly negative $G_{15}$ value in pseudo-OBC indicates a weak cross-link due to non-zero $R_{BC} = 2.95$ k$\Omega$. By floating the inner electrode of the contact arm BC, we can achieve PBC with $G_{15} \approx$ -1, as shown in the colormap of matrix $G$ in Fig. 3C. The deviation from the ideal value of $G_{15} = -1$ can be attributed to the calibration issues as was discussed previously. Figure 3D shows a smooth and continuous transition of $G_{15}$ from the pseudo-OBC to PBC as the $R_{BC}$ varies from $R_{BC} = R_{fridge}$ to $R_{BC} = \infty$, respectively. Furthermore, the eigenvalues of the G matrices corresponding to the PBC spread around a circle in the complex plane. As we gradually move toward pseudo-OBC, the radius of the eigenvalue circle becomes progressively smaller, and it eventually converges toward the real axis for OBC, as shown in Fig. 3E, which is consistent with the existing theoretical calculations (*52*).

We now explore the existence of the NHSE by determining the sum of probability densities, SPD, calculated from the eigenstates of the $G$ matrices. SPD at the site $i$ of the HN chain is defined as

$$SPD(i) = \sum_j |V_i^j|^2, \qquad (2)$$

where $V_i^j$ is the $i^{th}$ component of the $j^{th}$ normalized eigenvector, $\mathbf{V}^j$, of a $G$ matrix (*50*). Figure 3F shows the SPD as a function of the chain site index. We observe that, under pseudo-OBC, the SPD decays exponentially with the site index, whereas under PBC it remains approximately constant across all sites. This behavior indicates that the eigenstates are exponentially localized at the right-most site (contact arm 5) for pseudo-OBC, which is a hallmark of the NSHE. Alternatively, NSHE can be realized from the discrete time evolution of the eigenstates of the $G$ matrices. The time evolution operator, $U(\Delta t)$, associated with the Hamiltonian, $H_{HN} = G - E\mathbb{1}$, where E is the diagonal term of the $G$ matrix, is given by

$$U(\Delta t) = \exp\left(-\frac{iH_{HN}\Delta t}{\hbar}\right) = 1 - \frac{i\Delta t}{\hbar}H_{HN} + \cdots, \quad (3)$$

and the corresponding eigenvectors evolve as $|\Psi(\Delta t)\rangle = U(\Delta t)|\Psi(0)\rangle$, where $|\Psi(0)\rangle = \sum_n c_n |n\rangle_R$, and $c_n = {}_L\langle n|\Psi(0)\rangle$. $|n\rangle_R$ and $|n\rangle_L$ are the left and right eigenstates of $H_{HN}$ that satisfy ${}_L\langle m|n\rangle_R = \delta_{mn}$. For a finite-sized HN chain in real space, $H_{HN}$ takes the form

$$H_{HN} = \begin{pmatrix} 0 & J_{\text{left}} & 0 & 0 & 0 \\ J_{\text{right}} & 0 & J_{\text{left}} & \ddots & 0 \\ 0 & J_{\text{right}} & 0 & \ddots & 0 \\ 0 & \ddots & \ddots & \ddots & J_{\text{left}} \\ 0 & 0 & 0 & J_{\text{right}} & 0 \end{pmatrix}. \quad (4)$$

After $N$ time steps, the eigenvectors become

$$|\Psi(N\Delta t)\rangle \approx \left(1 - \frac{i\Delta t}{\hbar}H_{HN}\right)^N |\Psi(0)\rangle, \quad (5)$$

and the probability density, $|\Psi(N\Delta t)|^2$, converges to one end for a large enough $N$ for any given initial state $|\Psi(0)\rangle$ (see numerical simulation in the SI). This behavior can be intuitively understood by considering a random walk with biased transition probabilities, where the likelihood of jumping left exceeds that of jumping right, or vice versa (*69*). Regardless of the initial condition, the bias leads to a net drift toward the favored direction over time. In continuum models, this effect becomes particularly evident in the limit of extreme asymmetry, where the transition amplitude in one direction vanishes entirely. In that case, the dynamics are governed by a first-order differential equation, whose solutions describe unidirectional wave-like propagation (*70*).

Next, we explore the non-Hermitian dynamics in the metallic phase of the magnetic TI, away from the CNP, where the dissipationless chiral edge state is no longer present. Figure 4A shows the longitudinal and transverse resistance versus the back gate voltage $V_g$. We observe a deviation from resistance quantization away from the CNP ($V_g = -80$ V). Owing to the paraelectric nature of the STO substrate used for back-gating, the CNP shifts slightly away from $V_g = 4$ V during the back-gate sweep. Following the same procedure described earlier, we reconstruct the $G$ matrices for pseudo-OBC and PBC at $V_g = -80$ V. Figures 4(B-C) depict the resulting $G$ color maps. We observe finite coupling terms populating previously zero elements of the $G$ matrices, consistent with the absence of unidirectional transport. Instead, the topological surface states dominate transport at this gate voltage. In this regime, we observe that the lower diagonal elements are reduced by approximately half, while the upper diagonal elements range from $-0.2 e^2/h$ to $-0.3 e^2/h$, indicating asymmetric, bidirectional coupling between adjacent lattice sites. To illustrate this point, Fig. 4D presents the evolution of $G_{15}$ with back gate voltage. For pseudo-OBC, $G_{15}$ remains approximately $-0.12\ e^2/h$, whereas under PBC, it changes from $-1.1\ e^2/h$ at the CNP to $-0.56\ e^2/h$ at $V_g = -80$ V. This behavior establishes a direct link between non-Hermitian dynamics and non-reciprocal charge transport, providing a robust framework for understanding asymmetric transport phenomena in TIs. Furthermore, the eigenvalues evolve with the back gate voltage in pseudo-OBC, as shown in Fig. 4E, which exhibits a transition from circular shape in the complex plane, to approximately a line on the real axis as $V_g$ becomes more negative. The change in the shape formed by the eigenvalues is more pronounced at lower $V_g$'s.

In the HN model, the non-Hermicity is realized when the hopping amplitudes between the neighboring sites are direction dependent. Accordingly, the asymmetric, bidirectional coupling between the adjacent chain sites remains consistent with the HN model. The presence of non-reciprocal transport in our devices is further corroborated by second harmonic measurements of the longitudinal resistance, $R_{xx}$, in a Hall bar device fabricated on a separately grown magnetically doped TI (see Fig. S2 in the SI). Skew scattering between the chiral edge state and the Dirac surface states has been proposed as a possible mechanism for the non-reciprocity (*54*).

**CONCLUSION**
In conclusion, we have implemented the HN model in a magnetic TI and investigated the dynamics of the resulting non-Hermitian system. We find that due to the dissipationless nature of the chiral edge state at the CNP, we can obtain non-Hermitian $G$ matrices, exhibiting NHSE where the eigenstates are exponentially localized at one extremity of the HN chain in the pseudo-OBC case, while they are uniformly distributed along the chain in the PBC case. Away from the CNP, we observe asymmetric, bidirectional coupling between adjacent sites of the chain, where the Dirac surface states come into play, giving rise to non-reciprocity that is captured by the non-diminishing NHSE. Recent studies suggest that transport phenomena in the QAH phase are more complex than previously understood, revealing gate-tunable incompressible bulk strips that contribute to charge transport alongside the edge (*71*). Notably, edge currents can persist even after the QAH state breaks down (*72*). Our findings add to this ongoing exploration of bulk-boundary correspondence, non-Hermitian topological phases, and their implications for quantum transport in magnetic TIs.

**MATERIALS AND METHODS**
**Material Synthesis:** STO (111) substrate was prepared by a 1.5 hr soak in deionized water at 80 °C followed by a 3 hr annealing at 982 °C in a tube furnace with flowing oxygen gas. The substrate was then transferred into an ultra-high vacuum (approximately 2E-10 Torr) VEECO 620 molecular beam epitaxy chamber. The substrate was heated to 630 °C for 1 hour to outgas. Afterwards, the substrate was cooled to 255 °C and exposed to a Te overflux for 15 s to create a Te-rich environment for growth. A 3 QL Cr-(Bi, Sb)$_2$Te$_3$ was grown at a rate of approximately 0.25 QL/min. The Cr shutter was then closed to grow a 5 QL (Bi, Sb)$_2$Te$_3$ at the same growth rate. Finally, the Cr shutter was re-opened to deposit 3 QL Cr-(Bi, Sb)$_2$Te$_3$. The cell temperatures were controlled to achieve the desired beam equivalent pressure ratios (BEPR) between high-purity Cr (5 N), Bi (5 N), Sb (6 N), and Te (6 N) Knudsen effusion cells. The Te/(Bi+Sb) BEPR was approximately 17 to mitigate the formation of Te vacancies. The BEPR of Sb/Bi was approximately 1.2 to achieve an intrinsic charge neutrality point. The nominal Cr BEP was 4.7E-10 Torr.

**Device Fabrication:** A multi-terminal Corbino ring was first defined on the magnetic TI film with Heidelberg MLA150 maskless aligner. The magnetic TI film was subsequently etched away in a mixture of $H_2O_2$ (30%) / $H_3PO_4$ (85%) / $H_2O$ with a ratio by volume of 1:1:20 for 20 s at room temperature. Another photolithographic step was then performed to define the contacts. Finally, Cr/Au (10/50 nm) were deposited onto the sample using electron beam evaporator. Indium was applied to the backside of the substrate after device fabrication to create the back-gate electrode.

**Electrical Transport:** All the electrical transport measurements were performed inside a Bluefors LD250 dilution fridge, equipped with QDevil filtered DC lines, at the base temperature of approximately 12 mK, unless specified otherwise. The magnetic TI film was magnetized by applying a perpendicular magnetic field up to 0.5 T at 12 mK. To accurately measure the voltage, $V$, at each contact arm, the current, $I$, flowing into the inner electrode of the contact arm was measured. The voltage across the fridge line was subsequently subtracted from the measured voltage. The resistance was then calculated from $R = (V - R_{\text{fridge}}I)/I_{\text{ac}}$. The current was probed using a low-noise current-to-voltage pre-amplifier and a lock-in amplifier.

**ACKNOWLEDGEMENTS**
This research was supported in part by the Pennsylvania State University Materials Research Science and Engineering Center supported under the US National Science Foundation Grant No. DMR-2011839. Sample synthesis was supported by the Penn State Two-Dimensional Crystal Consortium-Materials Innovation Platform (2DCC-MIP) under NSF Grant No. DMR-2039351. R.E and S.K.O acknowledge support from Air Force Office of Scientific Research (AFOSR) Multidisciplinary University Research Initiative (MURI) Award No. FA9550-21-1-0202.

**Figures**

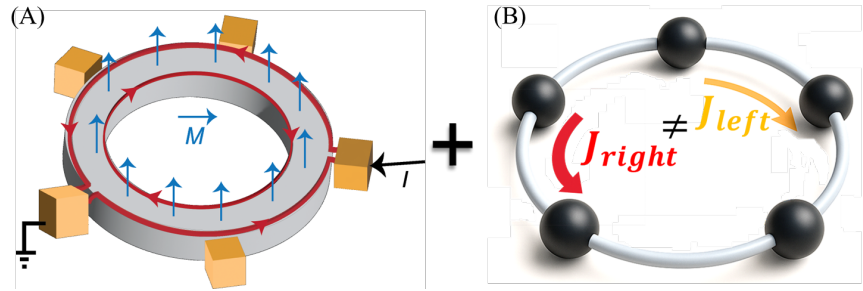

**FIG. 1. Material and geometric topologies.** (A) A schematic of the quantum anomalous Hall effect in a Corbino geometry. The magnetization direction within the TI is pointing up. The dissipationless chiral edge current corresponding to Chern number $C = 1$ flows counter-clockwise along the outer perimeter, and clockwise along the inner perimeter. The current injected into one of the terminals will flow counter-clockwise to ground. (B) A simple five-site Hatano-Nelson model with an open boundary condition. The hopping amplitudes $J_{left}$ and $J_{right}$ are unequal, resulting in asymmetric, bidirectional coupling between the adjacent sites. In the case of the QAH insulator, the hopping is unidirectional, i.e., $J_{left} = 0$ and $J_{right} = 1$.

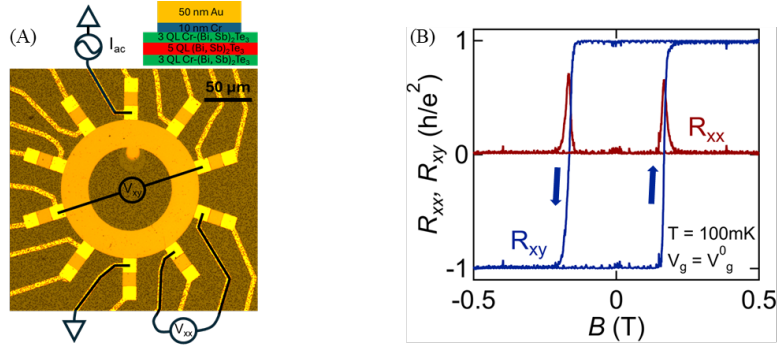

**FIG. 2. Quantum anomalous Hall Insulator characterization.** (A) Optical image of the device showing the Corbino geometry with 5 pairs of contact arms. The configuration for current injection and voltage probes used to measure the four-terminal longitudinal and transverse resistances is indicated. A schematic cross-section of the device is shown at the top. (B) The longitudinal, $R_{xx}$, and Hall, $R_{xy}$, resistance measured at the CNP ($V_g^0 = 4$ V) at 100 mK. Arrows indicate the directions of the magnetic field sweep.

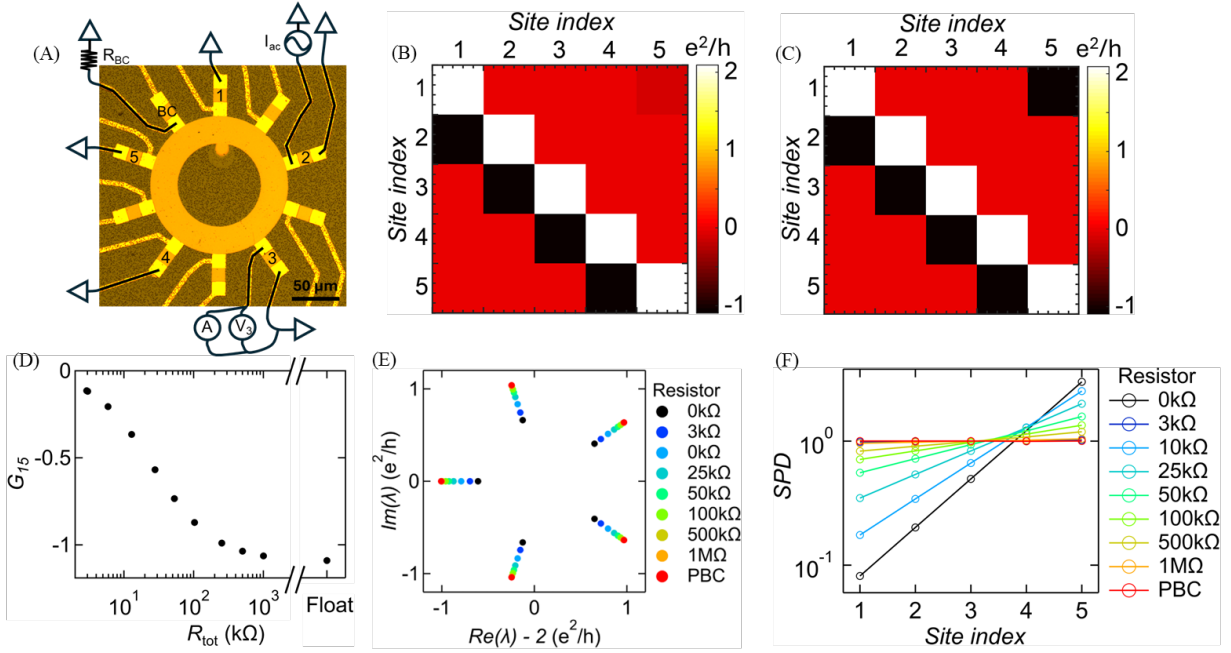

**FIG. 3. Non-Hermitian dynamics in the quantum anomalous Hall regime.** (A) Optical image of the device and measurement setup for the resistance matrix element $R_{32}$. Contact arms used for current/voltage probes are labeled 1-5 in the clockwise direction. A variable resistor $R_{BC}$ is connected from the inner electrode of contact arm BC to ground, enabling continuous tuning of the circuit boundary condition. Due to the fridge line resistance, the effective resistance between the inner electrode and ground is $R_{tot} = R_{BC} + R_{fridge}$, where $R_{fridge}$ = 2.93 k$\Omega$. (B, C) Colormaps of the $G$ matrices corresponding to pseudo-OBC and PBC, respectively. Site index corresponds to the contact arm labels. The matrix elements are normalized to $e^2/h$. (D) Conductance matrix element $G_{15}$ versus $R_{tot}$. "Float" denotes $R_{BC} = \infty$. (E) Eigenvalues of the conductance matrices for different values of $R_{BC}$, plotted in the complex plane. The real parts of the eigenvalues are shifted by $2e^2/h$, centering the distribution at the origin. In this measurement, PBC corresponds to $R_{BC} = \infty$. (F) Sum of the probability density (*SPD*) plotted against site index. During the transition from OBC to PBC, the *SPD* evolves from exponential localization at a single site to uniform distribution across the sites, a hallmark of the non-Hermitian skin effect.

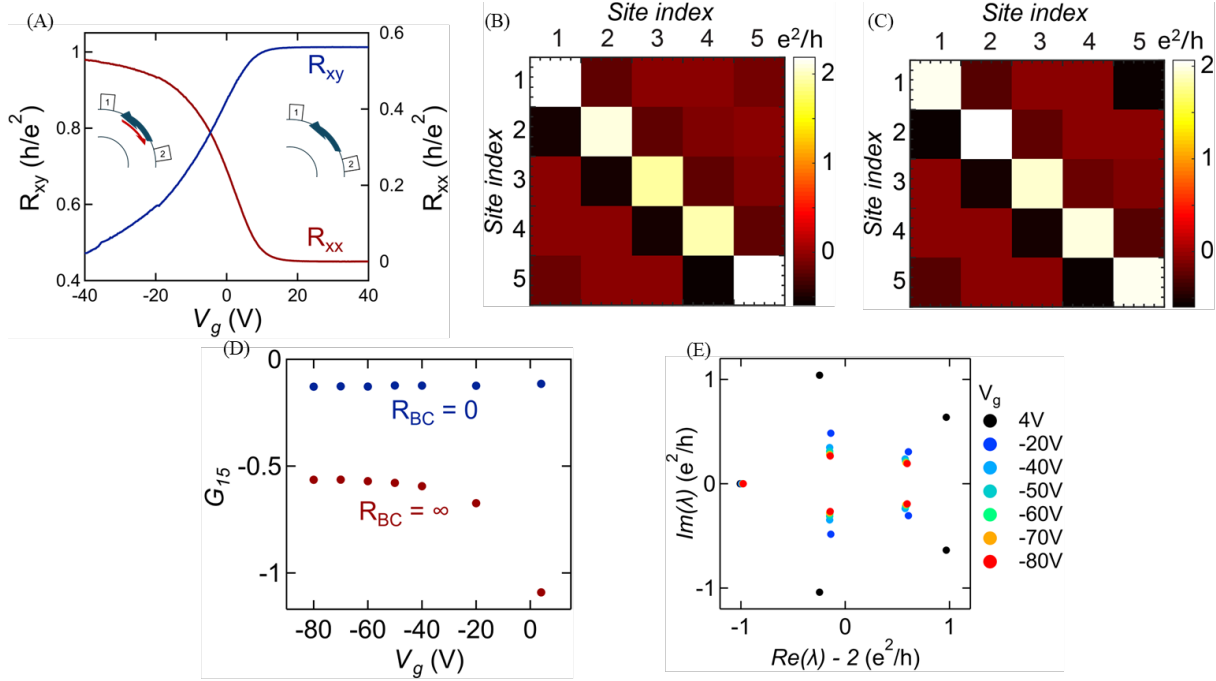

**FIG. 4. Non-Hermitian dynamics outside of the quantum anomalous Hall regime.** (A) Gate dependence of the longitudinal ($R_{xx}$) and Hall ($R_{xy}$) resistances. A dissipationless chiral edge state exists at the CNP, as depicted in the right inset for electrodes 1 and 2. As the gate is tuned away from the CNP, bidirectional edge states with asymmetric coupling emerge, illustrated in the left inset for the same electrodes. Due to the paraelectric nature of the STO substrate, the CNP is slightly shifted from $V_g$ = 4 V. (B, C) Colormaps of the $G$ matrices corresponding to pseudo-OBC (blue dots) and PBC (red dots) at $V_g$ = -80 V. (D) Gate dependence of the conductance matrix element $G_{15}$ for pseudo-OBC and PBC. (E) Gate dependence of the eigenvalues of the conductance matrices in pseudo-OBC. Away from the CNP, the circular distribution of eigenvalues in the complex plane progressively collapses toward the real axis.

# Supplementary Materials for
# Non-Hermitian Topology in Quantum Anomalous Hall Insulators

Le Yi[1], Emma Steinebronn[1], Nitin Samarth[1,2], Ramy El-Ganainy[3], Şahin L. Özdemir[3], and Morteza Kayyalha[4, *]

[1]Department of Physics, The Pennsylvania State University, University Park, Pennsylvania 16802, USA
[2]Department of Materials Science and Engineering, The Pennsylvania State University, University Park, Pennsylvania 16802, USA
[3]Department of Electrical and Computer Engineering, Saint Louis University, St. Louis, Missouri 63103, USA
[4]Department of Electrical Engineering, The Pennsylvania State University, University Park, Pennsylvania 16802, USA

## S1. Conductance matrices with explicit values

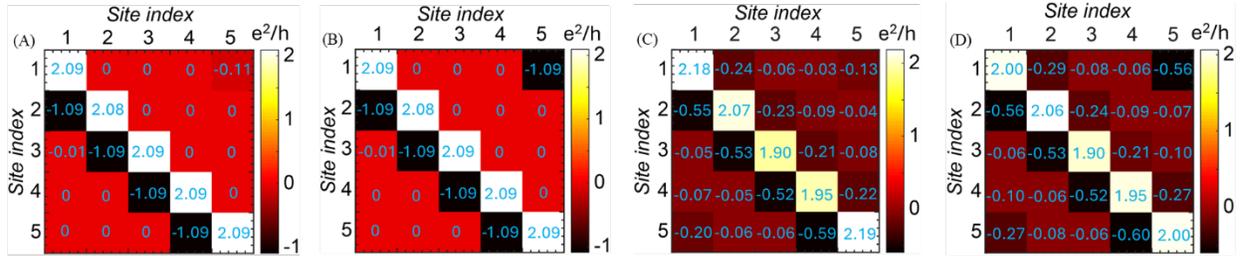

FIG. S1. **Conduction matrices.** Colormaps of the G matrices associated with the (A) pseudo-OBC and (B) PBC at $V_g = 4\ V$, and the (C) pseudo-OBC and (D) PBC at $V_g = -80\ V$, with the values of the matrix elements shown.

## S2. Second harmonics in magnetically doped TI

The longitudinal voltage drop, $V_{xx}$, up to the second order is described by
$$V_{xx}/l = R_0 i - \gamma R_0 c \widehat{M} i^2,$$
where $l$ is the distance between the contact pair, $R_0$ is the linear resistance, $\gamma$ is the strength of non-reciprocity, $c = +/-1$ for the contact pair at the top and bottom of the Hall bar, respectively, and $\widehat{M} = +/-1$ for upward and downward magnetization, respectively (*1*). Inversion symmetry breaking leads to oppositely signed non-reciprocal resistances between the two opposite edges (*1*). The direction of the magnetization depends on the applied external magnetic field. Thus, the non-reciprocal resistance switches the sign while crossing the zero-field point.

The Hall bar structure, as shown in Fig. S2(A), was fabricated on a magnetically doped TI that was grown separately from the film reported in the main text. For the second harmonic measurement, the longitudinal resistance, $R_{xx}$, is measured along both the top and bottom edges of the Hall bar device at $V_g = 48.5$ V under the temperature of $T = 1.2$ K. An AC excitation of $I_{ac} = 2$ μA is injected into the source of the Hall bar with a lock-in amplifier, and the resulting second harmonic voltage, which is then converted to resistance, $R_{xx}^{2\omega}$, is simultaneously captured and shown in Fig. S2(B) and (C) for oppositely swept magnetic field directions.

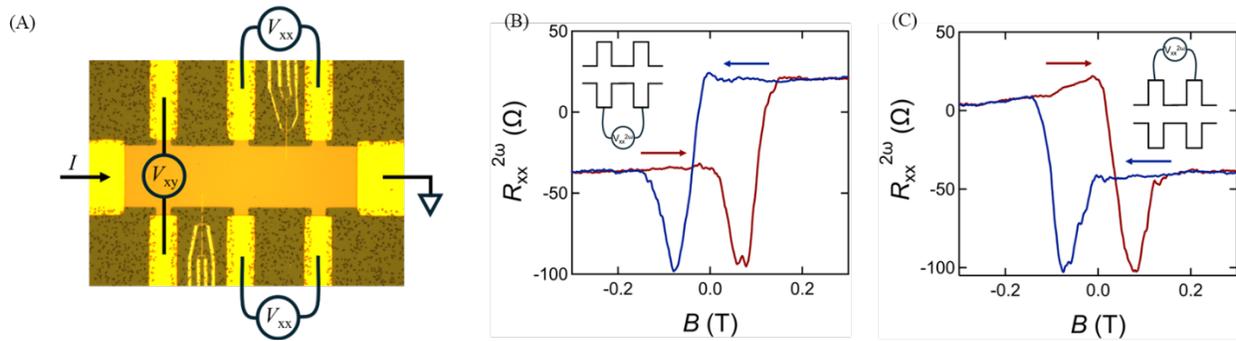

FIG. S2. **Second-harmonic measurement.** (A) Optical image of the Hall bar device used for second harmonics measurement. The Cr/Au (5/50 nm) contacts are deposited on top of the etched Cr-doped TI. The Hall measurement setup is also shown. (B) and (C) show the magnetic field dependences of the second harmonic signals for the longitudinal resistances measured from the top (B) and bottom (C) of the Hall bar as shown in the insets, which correspond to the setup in (A). Blue and red curves correspond to magnetic field sweeps in different directions, as indicated by arrows.

## S3. Matlab code for numerical simulation

The following Matlab code simulates non-Hermitian skin effect of the eigenvector, $|\Psi(N\Delta t)\rangle$, after n iterations, as described by Eqn. (5) in the main text.

```matlab
% Initialize the left and right hopping amplitudes
% The NHSE is more prominent as the difference between J_left and
% J_right increases
J_left = 0;
J_right = 1;

% Create a 5 by 5 matrix
N = 5;
H = zeros(N,N);

% Populate the off diagonal terms with J_left and J_right
for n = 1:N-1
   H(n,n+1) = J_left;
    H(n+1,n) = -J_right;
end

dt = 0.1;         % Time step size
In = rand(N,1);   % Randomly select an initial state
% For 100 time step iterations
for n = 1:100
   Out = (eye(N,N) - 1i*dt*H)^n*In;  % Implement eqn. (5) in the main text
end
```

```matlab
Out = abs(Out) / max(abs(Out))      % Normalize the output state
```